\documentclass[10pt,onecolumn]{article}
\usepackage[dvips]{graphicx}
\usepackage{multicol}
\newcounter{no_float_fig}
\begin{document}
\title{Automatic Quantum
Error Correction.}
\author{Jeff P Barnes and Warren S Warren \\
Department of Chemistry, Princeton University, \\
Princeton, NJ 08544-1009 }
\date{ (\today) }
\maketitle
\begin{abstract}
Criteria are given by which dissipative evolution
can transfer populations and coherences between
quantum subspaces, without a loss of coherence.
This results in a form of quantum error correction
that is implemented by the joint evolution of a system
and a cold bath.   It requires no external intervention
and, in principal, no ancilla.   
An example of a system that protects a qubit
against spin-flip errors is proposed.
It consists of three spin 1/2 magnetic particles,
and three modes of a resonator.   The qubit is the
triple quantum coherence of the spins, and the
photons act as ancilla.   This article is a greatly
expanded version of a letter submitted to 
{\it Physical Review Letters}.
\end{abstract}
PACS number(s): 03.67.Lx
%
%
\setcounter{page}{1}
\section{Introduction}
Quantum computation is of interest because
algorithms have been discovered with a significant
speed-up over any classical algorithm \cite{shor,grover},
although these may be unique cases \cite{beals}.
It is very likely that any physical implementation of
a quantum computation will require some form of active
quantum error correction.   Quantum error correcting codes
(QECC) have been devised \cite{shor,steane,qecc5,qecc7,preskill} 
and experimentally demonstrated \cite{qecnmr} 
that can protect a set of states, the codewords, 
against a set of errors.   QECC is similar in spirit
to quantum erasure experiments \cite{scully}, 
but with the twist that one is not allowed
to manipulate the environment.   The surprising fact is
that one can still disentangle the codewords from the
environment, by transferring the entanglement to another
set of states, the ancilla.
\par
However, implementing QECC is a formidable task.
There is a high premium placed on using as few qubits
(two-level systems) as possible, 
because as quantum systems grow in size,  
the number of transitions to be manipulated, unwanted
thermal effects \cite{warren}, and decoherence rates
\cite{giulini} all increase exponentially.
But, to take a specific example, the fault-tolerant
error detection and repair of even a single qubit
can require 15 physical qubits, 5 to store the
two codewords, and 10 of which must be in 
known states of zero entropy \cite{qecc5}.
In addition, 
depending upon how one counts a ``logic gate'',
as many as 28 coherent manipulations of
pairs of qubits are required for each 
repair, because 
``measuring the stabilizer'' means finding 
the eigenvalues of operators such as 
$I_{x1}I_{x2}I_{z3}I_{z5}$
(see Fig. (2) of Ref. \cite{qecc5}).
Such control over a 32,768-level system 
is a daunting task, even for a highly 
coherent spectroscopy such as NMR.
Although the efficiency of QECC improves for
larger computations, a physical scale-up factor 
of 22 is still required to factorize a 
thousand-digit number \cite{steane2}.
Part of the difficulty stems
from the need to know which error 
has struck, in order to repair it.
This is because different errors 
rotate the codeword states about 
separate axes in Hilbert space.
By containing information
about which error occurred, 
the ancilla also provide a 
conditional axis about which rotation
can coherently repair an error.   
Although this seems like an air-tight
argument, there is another way to approach
QECC, which we will explore here.
\par
To begin with, note how curious it is 
that QECC can assign a 
unique status to the codewords.   
While a classical probability 
space inherently contains a privileged basis, 
Hilbert space does not, and this difference 
has some striking consequences \cite{bell}.
In order to function, QECC requires access to ancilla
in a state of zero entropy \cite{nielsen}, which
suggests that one could view QECC as a controlled
cooling of the system.   It is dissipative evolution 
that adds classical aspects back into Hilbert space.
A large body of work exists that model a diverse range 
of relaxation phenomena in magnetic \cite{slichter} 
and optical resonance \cite{walls,zubairy}.   
They use Lindblad equations of motion \cite{giulini}.
An earlier approach that did use a Lindblad equation
\cite{paz} implemented QECC as a limit 
of very fast external manipulation.
In contrast, we seek an approach that is distinct
from the concepts of error detection and repair.
\par
We show here that dissipation can be used to implement
an ``automatic quantum error correction'' (AQEC),
so called because error correction results exclusively
from the joint evolution of a system coupled to a cold,
Markovian bath.   No intervention is required by the
programmer, and in theory, no ancilla are required,
although this would be unlikely in practice.
Clearly, dissipation can be used to stabilize two 
distinct states of a quantum system that could
store a classical bit of information.   
What is not obvious, is 
whether such a system could also
hold a qubit, since dissipation 
usually destroys coherence.
The key ideas are to use codewords such that
errors must add energy to the system, and
to set up the evolution of the system such
that excitation and environmental entanglements
are expelled from distinct codewords in a
symmetric way.   This prevents the bath
from gaining information on the codewords,
and thus coherence can be maintained.
In the last section, we outline a system
that utilizes three magnetic spin 1/2 particles,
and three photons, to implement an AQEC that
protects against spin-flip errors.
It requires only well understood interactions
from magnetic resonance spectroscopy, and
is intended to show that AQEC has potentially
real-world applications.
\section{A simple QECC example.}
To begin, let us review the idea of
quantum error correction by way of
a simple method that protects a single 
quantum state against 
environmental entanglements \cite{knill}.
The idea is similar to that of a 
quantum eraser experiment \cite{scully},
but with the twist that one is not allowed to 
interact the environment.   
Three two-level systems, labeled as $S$,
$A$ and $E$, are initially in the state
$( a|1_S \rangle + b |0_S \rangle )
|0_E \rangle |0_A \rangle$.
The goal is to keep $S$ in its current state.
An interaction between $S$ and $E$ creates
the new state $( a |1_S\rangle |p_E\rangle
+ b |0_S \rangle | q_E\rangle )|0_A\rangle$.
The environment is scattered 
into two states, $|p_E \rangle$ and $|q_E \rangle$.
When $|\langle p_E | q_E \rangle | < 1$,
the final state of $E$ depends upon the initial
state of $S$, so they are entangled.
If $| p_E\rangle = -|q_E\rangle$, the
phase of $S$ has been flipped.
To repair $S$, first note that the
entangled state can be written as:
\[
\frac{1}{2} \: \Bigg\{
\bigg( a|1_S \rangle + b|0_S \rangle \bigg)
\bigg( |p_E \rangle + |q_E \rangle \bigg)
+ \bigg( a|1_S \rangle - b|0_S \rangle \bigg)
\bigg( |p_E \rangle - |q_E \rangle \bigg)
\Bigg\} |0_A \rangle
\]
Suppose we can externally manipulate the qubits.
Conditionally flip $A$, if the sign of the
state $S$ is flipped from what we expect it
to be.   In the language of QECC, this is
``measuring the stabilizer'', or ``detecting
the error''.   $A$ serves as the memory.
Next, flip the sign of $S$, 
conditional on if $A$ detected an error.  
This is ``repairing the state''.
Both of these actions are unitary 
transforms on $S$ and $A$ only; the environment
is never directly manipulated.
After measuring the error, the new state is:
\[
\frac{1}{2} \Bigg\{
\bigg( a |1_S \rangle + b|0_S \rangle \bigg)
\bigg( |p_E \rangle + |q_E \rangle \bigg) |0_A \rangle
+ \bigg( a |1_S \rangle - b|0_S \rangle \bigg)
\bigg(|p_E \rangle - |q_E \rangle \bigg) |1_A \rangle
\Bigg\},
\]
and then, after the repair,
\[
\bigg( a|1_S \rangle + b|0_S \rangle \bigg)
\frac{1}{2} \Bigg\{
\bigg( |p_E \rangle + |q_E \rangle \bigg) |0_A \rangle
+ \bigg( |p_E \rangle - |q_E \rangle \bigg) |1_A \rangle
\Bigg\}
\]
The original state of $S$ has re-emerged!
The entanglement between $S$ and $E$ was 
transferred to be between $A$ and $E$,
without ever touching $E$.   In order for this
scheme to work, it is crucial that $A$ is initially
in a single pure state, or in a state of zero entropy.
We expect that we can achieve this by cooling $A$
down to 0 $^\circ$K by the third law of thermodynamics.
However, there are systems such as protons in ice
or frustrated spin lattices \cite{ramirez} that
are postulated to violate the third law.
Since cooling these systems still leaves them
in a state of non-zero entropy, one should avoid
using them as ancilla.
\subsection{A dynamical re-formulation of the above example.}
The next step is to transform the
above error correcting method into the
language of a quantum system, relaxing
towards equilibrium.
We will make use of the 
operator formalism of NMR \cite{ernst}.
The qubit states are $|0\rangle$ and $|1\rangle$,
and the projection operators are
$I_\alpha = |1\rangle \langle 1|$ and
$I_\beta = |0\rangle \langle 0|$.
The raising and lowering operators are
$I_+ = |1\rangle \langle 0|$ and
$I_- = |0\rangle \langle 1|$, respectively.
The Hermitian Pauli operators are
$I_x = (I_+ + I_-)/2$,
$I_y = (I_+ - I_-)/2i$, and
$I_z = (I_\alpha - I_\beta)/2$,
and $\vec{I}$ is the vector
formed by them.   The subscript
also indicates which spin is acted upon,
so $I_{n,x}$ acts only on spin $n$.
This section is similar to that of Ref. \cite{paz}, 
but with the difference that
the measurement of the syndrome 
and the repair process are treated more
explicitly.
\par
The Hamiltonian of Eq. (\ref{eqn:H1}).
is designed to continuously implement the
example of the last section.   To keep $S$ in 
the state $|1_S\rangle$, first flip $A$ at
the rate $d$, if $S$ departs from $|1_S\rangle$.
If $A$ has flipped, then $S$ is flipped at a rate $r$.   
To complete the process, $A$ is cooled at a rate
$c$ by interaction with a bath
of harmonic oscillators with a broad spectral
response.  If the bath temperature is low
in comparison to the separation of the levels
of $A$, then the density matrix $\rho$  
evolves as \cite{giulini,walls},
\begin{equation} \begin{array}{c}
H =  r ( I_{A,\beta} + I_{A,\alpha} I_{S,x} )
+ d ( I_{S,\alpha} + I_{S,\beta} I_{A,x} ) \\
\frac{\displaystyle \partial \rho }
{\displaystyle \partial t} =
-i [ H, \rho ] - c ( I_{A,\alpha} \rho + 
\rho I_{A,\alpha} - 2 I_{A,-} \, \rho \, I_{A,+} )
\end{array} \label{eqn:H1} \end{equation} 
We suppose the errors occur rapidly in comparison
to the system dyanmics, so they are modeled as
instantaneous transforms.   However, slower
interactions can also be corrected \cite{paz}.
\par
Dissipative evolution is usually handled in Liouville
space, where $\rho$ is a vector, and transformations
like $-i[H,\rho]$ are matrix-vector multiplications 
\cite{superoperator}.   These matrices are called
superoperators, since they operate on operators.
Eq. (\ref{eqn:H1}) becomes a set of linear
differential equations, $\dot{\rho}$ = 
$\Gamma \rho$.  The elements of $\Gamma$ are
then indexed by how they transform the populations 
and coherences of an orthonormal set that spans
the Hilbert space:  each element of $\Gamma$
transforms a $|j\rangle \langle k|$ to 
a $|n\rangle \langle m|$.
When $c=0$, the evolution is unitary, and $\Gamma$ 
has two kinds of eigenvalues:
$\lambda$ = 0, corresponding to populations of 
eigenstates of $H$, $|n\rangle \langle n|$, and 
$\lambda$ = $\pm i \beta$, corresponding to 
coherences $|n\rangle \langle m|$ and 
$|m\rangle \langle n|$.   Unfortunately, Liouville
space also increases the problem size:
$N$ qubits now require an evolution superoperator
with $4^N$ eigenstates.   
\par
When evolution is dissipative,
$\Gamma$ is not a symmetric matrix.   It can still
be written as the outer product of its right and left
eigenvectors, $\Gamma$ = 
$\sum \lambda_n \vec{r}_n \otimes \vec{l}_n$,
but in general the $\vec{r}_n$ are not orthogonal.
However, $\vec{l}_n \cdot \vec{r}_m$ = $\delta_{nm}$, 
which allows one to formally solve the
equation of motion for an operator as $\vec{x}(t)$ = 
$\sum \vec{r}_n ( \vec{l}_n \cdot \vec{x}(0) \, )
\exp(\lambda_n t)$.   The structure of Eq. (\ref{eqn:H1}) 
implies that $\Gamma$ conserves $\mbox{tr}(\rho(t)\,)$, 
but a pure state will not necessarily remain pure.
\par
Does the system of Eq. (\ref{eqn:H1}) work?
Fig. (\ref{fig:graph}) plots the real parts of the
eigenvalues of $\Gamma$ of Eq. (\ref{eqn:H1})
for various values of $d$, $r$ and $c$.
The only stable states of $\Gamma$ will
be those with $\Re e(\lambda_n)=0$, and there is 
only one such state: $|1_S 0_A \rangle \langle 1_S 0_A |$.
This is how dissipative evolution can confer 
a privileged status on a state.
\par
A numerical integration of the system dynamics
also shows this.   Fig. (\ref{fig:graph2}) plots 
the linear entropy, defined as
$0 \le \mbox{tr}(\rho(t)-\rho^2 (t) \, ) \le 1$,
of $\rho(t)$, starting from the corrupted state 
$|0_S 0_A \rangle \langle 0_S 0_A |$.
The rate at which the error is repaired is dominated
by the eigenvalue of $\Gamma$ with the least negative,
but nonzero, real part.  Curiously, a larger $c$ 
is counterproductive, as it traps the state 
into a cycle:
\[ |0_S 0_A \rangle 
\begin{array}{c} \mbox{detect}\rightarrow \\
\leftarrow \mbox{cool} \end{array}
|0_S 1_A \rangle \;\; \mbox{repair} \rightarrow 
\;\; |1_S 1_A \rangle \;\; 
\mbox{cool} \rightarrow \;\;
|1_S 0_A \rangle
\]
\par
\includegraphics[angle=0,width=3in,height=2.5in]
{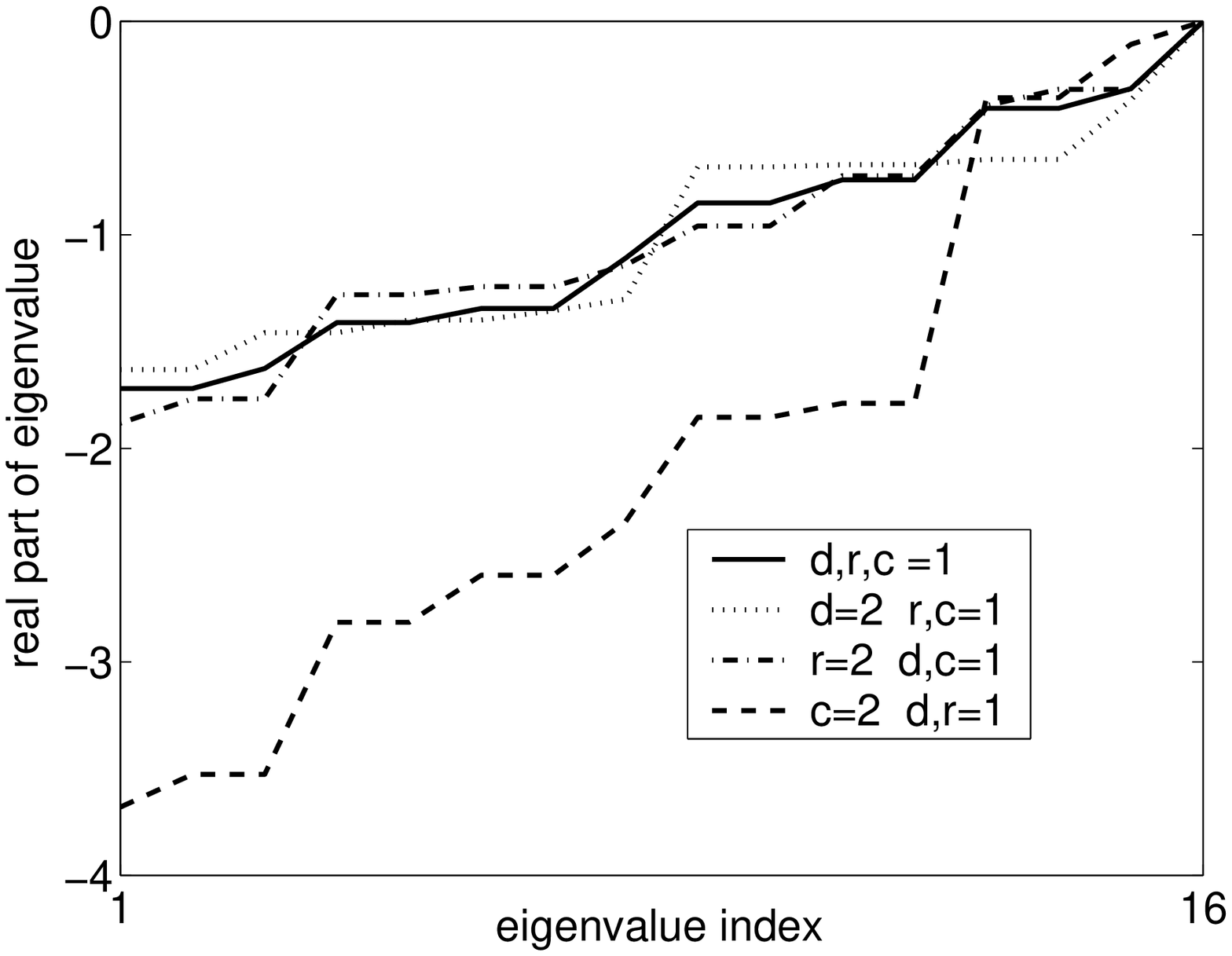}
\hfill
\includegraphics[angle=0,width=3in,height=2.5in]
{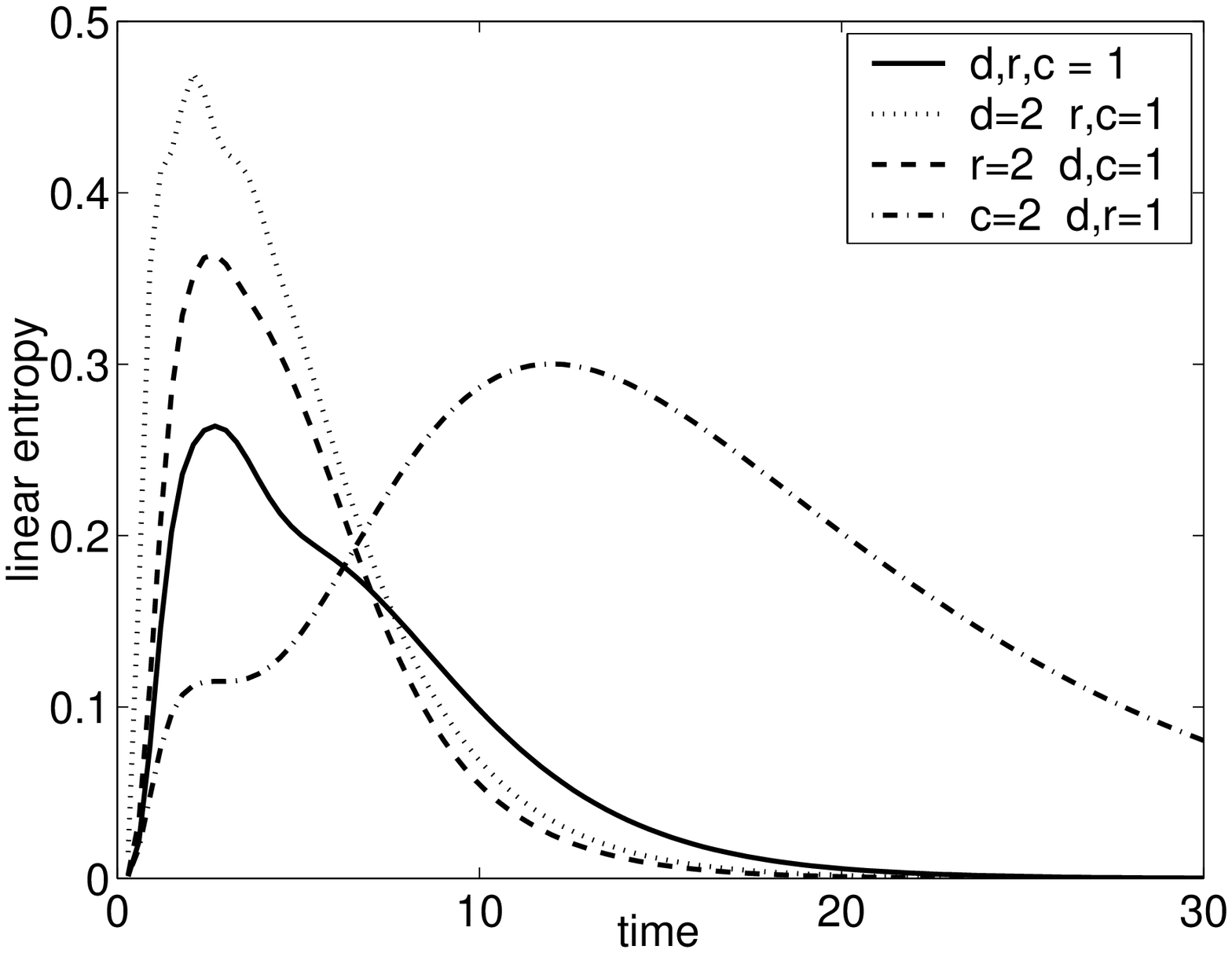}
\par
\parbox{3.0in}{
\refstepcounter{no_float_fig} \label{fig:graph}
\textbf{Fig. \ref{fig:graph}.}
The real parts of the 16 eigenvalues 
of the superoperator $\Gamma$
for some values of $d$, $r$ and $c$.
Note that there is only a single stable state.}
\hfill
\parbox{3in}{
\refstepcounter{no_float_fig} \label{fig:graph2}
\textbf{Fig. \ref{fig:graph2}.}  
The linear entropy
during the continuous error correction,
for $d$, $r$, $c$ = 1, and doubling each
parameter separately.   The starting state
is $\rho = |0_S 0_A \rangle \langle 0_S 0_A |$.}
\par
\section{Conditions for AQEC.}
\subsection{Repairing the Populations.}
We now show what conditions
are necessary in order that
dissipative evolution can automatically
protect a subspace of codewords against a
given set of errors.
We first suppose that the system obeys
a Lindblad equation of motion.
In general, this is not a trivial assumption \cite{giulini}.
The most speculative condition in deriving
a Lindblad equation is that the system 
and the bath initially factorize.
Curiously, it can be justified here on the grounds
that a properly working error correction should drive
the system to this state.
A more troublesome condition is that if degenerate
transitions are coupled to the Markov bath,
they must couple to orthogonal bath modes \cite{cooling,vankampen}.
\par
Two conditions can be stated immediately.
We are assuming that evolution for a sufficient
time, $T$, can repair any error.
Thus, $\exp(\Gamma \, T)$ must be a
repair superoperator.   Necessary and sufficient
conditions for its existence are known from QECC \cite{knill}.
Under these conditions, the original codeword
populations and coherences might be transported
elsewhere in Hilbert space, but they are not
destroyed.   
The second condition is that the codewords must
be immune from the influence of the bath, or that
they form a decoherence free subspace with respect
to the system / bath coupling \cite{lidar}.
\par
An example is instructive.   Suppose that
we are interested in protecting a two-codeword
system against spin-flip errors.
The system is split into two groups of qubits, 
$S$ and $A$, where the $A$ are continuously cooled.
This is not necessary for the general argument, 
which can be formulated entirely in terms of the 
eigenvalues of $\Gamma$.
However, it simplifies the physical interpretation.
The system evolution is given by \cite{walls,zubairy}
\begin{equation}
\frac{\partial \rho}{\partial t} =
-i [ \, H, \rho \, ] - 
\sum_n^{\mbox{ancilla}}
c_n \bigg( I_{n,\alpha} \rho +
\rho I_{n,\alpha} - 2 I_{n,-}
\rho I_{n,+} \, \bigg)
\label{eqn:lindblad} \end{equation}
where $H$ acts on both the $S$ and $A$.
The second term irreversibly draws population
from the $|1\rangle$ states of the ancilla
spins, and places it in the $|0\rangle$ states.
Choosing codewords of the form 
$|\psi_n \rangle |0_A \, \rangle$, for which
(1) the $A$ are in their ground states, and 
(2) the codewords are eigenstates of $H$,
will satisfy the criteria for the 
decoherence-free subspace.
\par
The need for the QECC conditions 
can be seen as follows.  Suppose
we choose the two-$S$ states
$|00\rangle$ and $|11\rangle$ 
as the codewords.   But then the errors 
$I_{1,x}|00\rangle$ and $I_{2,x}|11\rangle$ 
both result in the same state, $|10\rangle$.
Under the Markov approximation, 
the system can not know 
which codeword was the original codeword, 
and so $\Gamma$ can not repair these errors.
But the three-$S$ states 
$|000\rangle$ and $|111\rangle$ will work,
since the spaces spanned by all
the errors acting on each codeword are now disjoint.
Thus, errors should transfer separate codewords
into disjoint subspaces.
\par
Now consider how to repair the codeword
populations. The set of errors acting on a
codeword, and all the further states that the
corrupted codeword evolves into under $\Gamma$,
form a subspace, as indicated in Fig.~(\ref{fig:example}).
Call this subspace the ``funnel'' associated
with the codeword, but excluding the codeword state
itself.   The name is suggestive of its role
in AQEC.   The QECC conditions already require the
initially excited states to be disjoint between
separate codewords.   Thus, if we add the
third condition that $\Gamma$ draws all
population from each funnel state into its associated
codeword, and transfers no amplitude between funnels,
then the codeword populations are repaired.
\par
\parbox[t]{3in}{
\includegraphics[angle=0,width=3in,height=2.5in]
{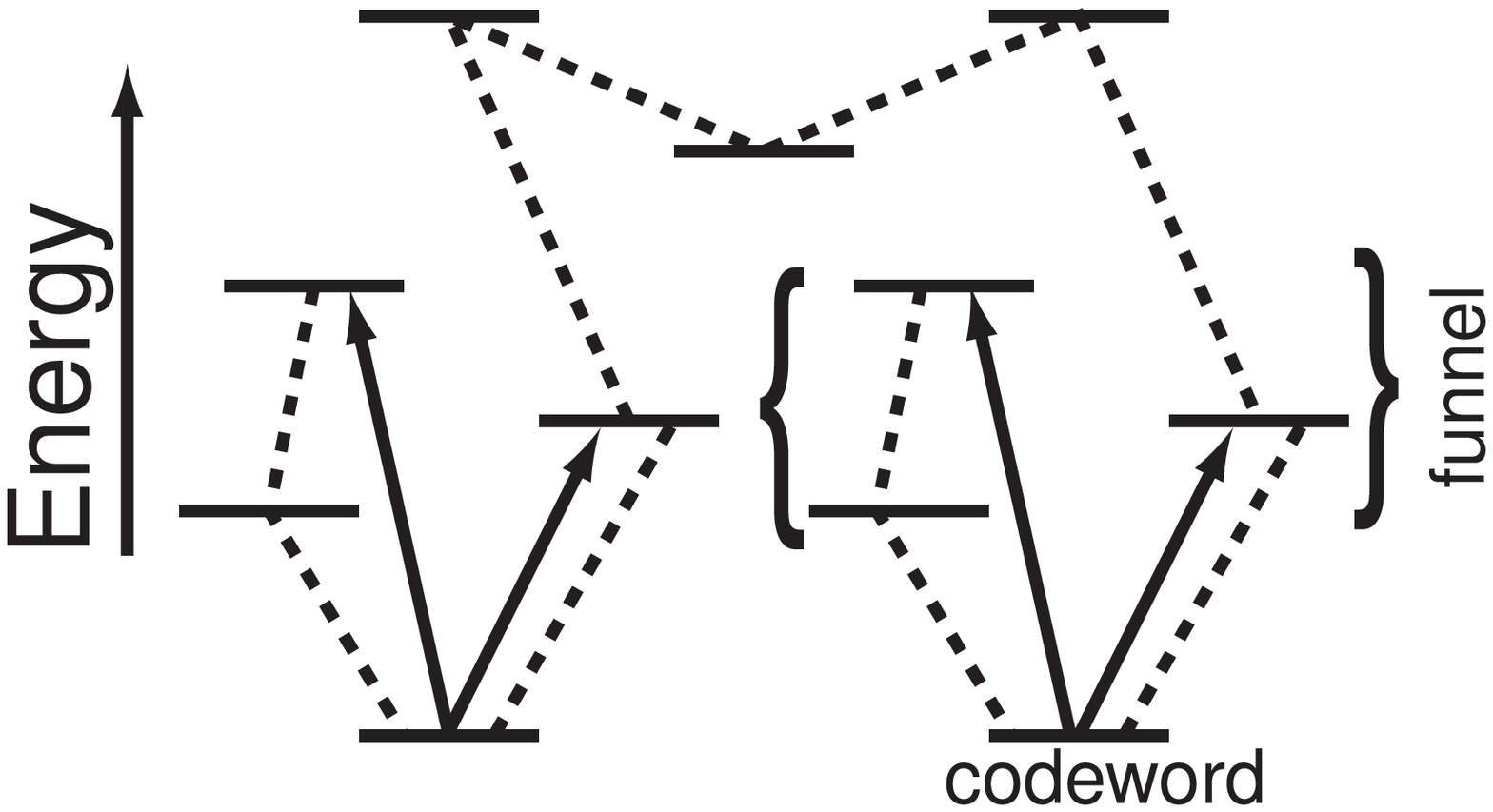}}
\hfill
\parbox[b][2.5in][c]{3in}{
\refstepcounter{no_float_fig} \label{fig:example}
\textbf{Fig. \ref{fig:example}.}  
The level diagram of a hypothetical
system.   The errors (solid arrows) transfer
amplitude into the disjoint funnels associated
with each codeword.  The three levels between
the brackets form the funnel for the labeled
codeword to the right.   Dissipative cooling
of selected transitions (dashed lines) then
returns the populations to their original
codewords.}
\par
An important difference between the usual
method by which QECC is implemented, and AQEC,
has emerged.   By placing the burden of the
repair on the system / bath coupling, AQEC,
in theory, requires no ancilla.   
Consider how to repair the error $I_{1,x}$,
acting on the three-$S$ and two-$A$ codewords
$|000,00\rangle$ and $|111,00\rangle$.
Suppose that the unnormalized states
$|100,00\rangle \pm |000,10\rangle$ and
$|011,00\rangle \pm |111,10\rangle$
are eigenstates of $H$.
The corrupted state $|100,00\rangle$
now periodically becomes the state
$|000,10\rangle$, where cooling of the
first ancilla returns it to the codeword $|000,00\rangle$.
However, we can not choose
$|010,00\rangle \pm |000,10\rangle$ and
$|101,00\rangle \pm |111,10\rangle$ as
eigenstates of $H$ in order to repair
the error $I_{2,x}$, because they are not
orthogonal to the first set.
It seems that we must choose
$|010,00\rangle\pm|000,01\rangle$ and
$|101,00\rangle\pm|111,01\rangle$ to
repair $I_{2,x}$, and
$|001,00\rangle\pm|000,11\rangle$ and
$|110,00\rangle\pm|111,11\rangle$ to
repair $I_{3,x}$.
In this strategy, we must be able
to distinguish between the different
errors in order to be able to repair
them, and counting $\epsilon$ different
errors requires log$_2\epsilon$ binary
digits (or ancilla qubits).
\par
But the third condition for AQEC only
requires that $H$ mix the states
$|100,00\rangle$, $|010,00\rangle$, and
$|001,00\rangle$ with $|000,10\rangle$,
even if only partially!   Because cooling
irreversibly draws probably away from the
excited ancilla, any degree of mixing will
do.   In other words, when the 4$\times$4
block of $H$ corresponding to the above
states is diagonalized, each eigenstate
should have a non-zero projection onto
the state $|000,10\rangle$, and similarly
for the other codeword.
We still require at least one ancilla here,
because the system was split into $S$ and
$A$, and only the $A$ are cooled.
If a bath / system coupling is found that
directly cools the funnel to codeword
transitions as in Fig.~(\ref{fig:example}),
then no ancilla are necessary.
However, not all the conditions for AQEC have
been stated yet.   The rest of these come
from the seemingly bizarre notion that we can use 
dissipation to restore a coherence.
\subsection{Repairing the Coherences.}
The QECC conditions ensure that the codeword
coherences are transferred, but not ``measured'',
by the environment.   AQEC must transfer them back.
What happens to coherences during dissipative
evolution is a subtle point, which is best explored
by way of a comprehensive example.
Codewords of the form $| \psi_n \rangle |00\rangle$,
with two ancilla, will serve this purpose.
The environment, $|e\rangle$, is initially
unentangled with the computer.  An interaction, $U$, 
can entangle the system so that \cite{barnes}
\begin{equation} \begin{array}{c}
U \bigg( a_0 | \psi_0 \rangle |00\rangle + 
a_1 | \psi_1 \rangle |00\rangle 
+ a_2 | \psi_2 \rangle |00\rangle 
+ a_3 | \psi_3 \rangle |00\rangle + \cdots \bigg)
| e \rangle = \\
\;\;\: a_0 \bigg(
u_0^{(0)} | \phi_0^{(0)} \rangle |00\rangle | e_0^{(0)} \rangle +
u_0^{(1)} | \phi_0^{(1)} \rangle |00\rangle | e_0^{(1)} \rangle + 
u_0^{(2)} | \phi_0^{(2)} \rangle |00\rangle | e_0^{(2)} \rangle + 
\cdots \bigg) \\
+ a_1 \bigg( 
u_1^{(0)} | \phi_1^{(0)} \rangle |00\rangle | e_1^{(0)} \rangle +
u_1^{(1)} | \phi_1^{(1)} \rangle |00\rangle | e_1^{(1)} \rangle +
u_1^{(2)} | \phi_1^{(2)} \rangle |00\rangle | e_1^{(2)} \rangle + 
\cdots \bigg) \\
+ a_2 \bigg(
u_2^{(0)} | \phi_2^{(0)} \rangle |00\rangle | e_2^{(0)} \rangle +
u_2^{(1)} | \phi_2^{(1)} \rangle |00\rangle | e_2^{(1)} \rangle +
u_2^{(2)} | \phi_2^{(2)} \rangle |00\rangle | e_2^{(2)} \rangle + 
\cdots \bigg) \\
+ a_3 \bigg(
u_3^{(0)} | \phi_3^{(0)} \rangle |00\rangle | e_3^{(0)} \rangle +
u_3^{(1)} | \phi_3^{(1)} \rangle |00\rangle | e_3^{(1)} \rangle +
u_3^{(2)} | \phi_3^{(2)} \rangle |00\rangle | e_3^{(2)} \rangle +
\cdots \bigg) \\
\end{array} \label{eqn:error} \end{equation}
After the error, the amplitude originally
in each codeword is spread throughout 
its funnel.   While the funnel states
$\{ |\phi_n^{(k)}\rangle \}$ can be chosen
as an orthogonal set for each $n$, this is
not true in general for the $\{ |e_n^{(k)}\rangle \}$.
\par
As yet, there is no constraint on either how separate
codewords can excite the ancilla qubits, or how the
dynamics of the repair should proceed.   Suppose
$H$ uses the first ancilla to repair the codewords
$n$ = 0 and 1, the second ancilla to repair 
$n$=2, and both ancilla to repair $n$=3.
That is, $H$ mixes each $|\phi_0^{(k)}\rangle|00\rangle$
with $|\psi_0\rangle|10\rangle$, and so on.
Let us follow an argument
analogous to the ``quantum jump'' 
approach \cite{zubairy}.
The relaxation process is divided up into small
time steps, $\Delta t$, during which the system 
and bath evolve separately.  
At the end of each interval, a fraction 
of the amplitude in each excited ancilla state
jumps into a de-excited state.
Tagging on two more qubits to represent two modes
of the cold bath, at the end of each time
interval, a fraction of the amplitudes make
the following jumps:
\[ \begin{array}{ccc}
|\psi_0\rangle |10\rangle |e_0^{(k)} \rangle |00\rangle &
\rightarrow &
|\psi_0\rangle |00\rangle |e_0^{(k)} \rangle |10\rangle \\
|\psi_1\rangle |10\rangle |e_1^{(k)} \rangle |00\rangle &
\rightarrow &
|\psi_1\rangle |00\rangle |e_1^{(k)} \rangle |10\rangle \\
|\psi_2\rangle |01\rangle |e_2^{(k)} \rangle |00\rangle &
\rightarrow &
|\psi_2\rangle |00\rangle |e_2^{(k)} \rangle |01\rangle \\
|\psi_3\rangle |11\rangle |e_3^{(k)} \rangle |00\rangle &
\rightarrow & \!\!\! \left\{
\begin{array}{c}
|\psi_3\rangle |10\rangle |e_3^{(k)} \rangle |01\rangle \\
|\psi_3\rangle |01\rangle |e_3^{(k)} \rangle |10\rangle \\
\end{array} \right. \\
|\psi_3\rangle |10\rangle |e_3^{(k)} \rangle |00\rangle &
\rightarrow &
|\psi_3\rangle |00\rangle |e_3^{(k)} \rangle |10\rangle \\
|\psi_3\rangle |01\rangle |e_3^{(k)} \rangle |00\rangle &
\rightarrow &
|\psi_3\rangle |00\rangle |e_3^{(k)} \rangle |01\rangle \\
\end{array} \]
The entire process is repeated until a time $T$, 
when the relaxation process is complete.
\par
The heart of the argument relies on the idea
that, in the limit of a large number of cold bath
modes interacting with the ancilla, it
is very likely that different ancilla that de-excite
at different times, will transfer their excitation
to orthogonal modes of the bath.   Once excited,
these modes do not further influence the evolution
of the computer, {\it i.e.} there is no back-reaction
from the bath.   In this case, after equilibrium
is reached, the final wavefunction is given by:
\begin{equation}
\begin{array}{c}
\;\;\: a_0 |\psi_0\rangle |00\rangle 
u_0^{(0)} |e_0^{(0)} \rangle \bigg(
c_0^{(0)} (\Delta t)|100000 \rangle + 
c_0^{(0)} (2\Delta t)|010000 \rangle +
c_0^{(0)} (3\Delta t)|001000 \rangle + \cdots \bigg) \\
+ a_0 | \psi_0 \rangle |00\rangle 
u_0^{(1)} |e_0^{(1)} \rangle \bigg(
c_0^{(1)} (\Delta t) |100000 \rangle +
c_0^{(1)} (2\Delta t) |010000 \rangle +
c_0^{(1)} (3\Delta t) |001000 \rangle + \cdots \bigg) \\
+ a_0 | \psi_0 \rangle |00\rangle 
u_0^{(2)} |e_0^{(2)} \rangle \bigg(
c_0^{(2)} (\Delta t) |100000 \rangle +
c_0^{(2)} (2\Delta t) |010000 \rangle +
c_0^{(2)} (3\Delta t) |001000 \rangle + \cdots \bigg) \\
\cdots \\
+ a_1 |\psi_1\rangle |00\rangle 
u_1^{(0)} |e_1^{(0)}  \rangle \bigg(
c_1^{(0)} (\Delta t)|100000 \rangle + 
c_1^{(0)} (2\Delta t)|010000 \rangle +
c_1^{(0)} (3\Delta t)|001000 \rangle + \cdots \bigg) \\
+ a_1 |\psi_1\rangle |00\rangle 
u_1^{(1)} |e_1^{(1)} \rangle \bigg(
c_1^{(1)} (\Delta t)|100000 \rangle + 
c_1^{(1)} (2\Delta t)|010000 \rangle +
c_1^{(1)} (3\Delta t)|001000 \rangle + \cdots \bigg) \\
\cdots \\
+ a_2 |\psi_2\rangle |00\rangle 
u_2^{(0)} |e_2^{(0)} \rangle \bigg( 
c_2^{(0)} (\Delta t)|000100 \rangle +
c_2^{(0)} (2\Delta t)|000010 \rangle +
c_2^{(0)} (3\Delta t)|000001 \rangle + \cdots \bigg) \\
+ a_2 |\psi_2\rangle |00\rangle
u_2^{(1)} |e_2^{(1)} \rangle \bigg( 
c_2^{(1)} (\Delta t)|000100 \rangle +
c_2^{(1)} (2\Delta t)|000010 \rangle +
c_2^{(1)}(3\Delta t)|000001 \rangle + \cdots \bigg) \\
\cdots \\
+ a_3 |\psi_3\rangle |00\rangle 
u_3^{(0)}  |e_3^{(0)}  \rangle \bigg( 
c_3^{(0)} (\Delta t,2\Delta t)|100010\rangle +
c_3^{(0)} (2\Delta t,\Delta t)|010100\rangle +
c_3^{(0)} (\Delta t,3\Delta t)|010001\rangle \\
\hspace{2cm} +
c_3^{(0)} (3\Delta t,\Delta t)|001100\rangle +
c_3^{(0)} (2\Delta t,3\Delta t)|010001\rangle +
c_3^{(0)} (3\Delta t,2\Delta t)|001010\rangle + \cdots \bigg) \\
\end{array} \label{eqn:jump} \end{equation}
For a funnel that uses a single ancilla,
the $c_n^{(k)}(m\Delta t)$ are the amplitude
to start in the state 
$|\phi_n^{(k)}\rangle|00\rangle|e_0^{(k)}\rangle$,
and transfer an excitation to the bath at $m\Delta t$.
Formally, it can be constructed from the system
propagator, $\exp(-i H m\Delta t / \hbar)$, and
matrix elements of the system / bath interaction.
Using more than one excited ancilla results
in a two-time dependence for the $c$.
All these functions approach zero for $t \rightarrow T$,
due to the irreversible loss of amplitude from the
funnel states at earlier times.
\par
The important point is that the $c_n^{(k)}(m\Delta t)$,
for different $n$ and $m$, uniquely label orthogonal
modes of the bath.    To see the consequences of this,
form $\rho$ from Eq. (\ref{eqn:jump}) by tracing out
the bath and environment.   The populations look
like this:
\begin{equation}
|\psi_0 \rangle |00\rangle \langle \psi_0 | \langle 00| \,
|a_0|^2 \times \sum_m
| u_0^{(0)} c_0^{(0)}(m\Delta t)|e_0^{(0)}\rangle +
u_0^{(1)} c_0^{(1)}(m\Delta t)|e_0^{(1)}\rangle + \cdots |^2
\label{eqn:popprod} \end{equation}
with similar expressions for the other codewords.
Because of the earlier conditions on $\Gamma$,
the populations must be repaired (there is no where
else for the populations to go).
Thus, the sum in Eq. (\ref{eqn:popprod}) is one.
However, from Eq. (\ref{eqn:jump}), it is easy
to see that the coherence $|\psi_2\rangle \langle \psi_0|$
is zero!  Using orthogonal ancilla states between
$n$=0 and 2 resulting in these codewords
becoming entangled with orthogonal bath modes.
What has happened, is that the 
pattern of excitation in the bath 
can be used to determine the probability 
to be in each codeword.   Using orthogonal
ancilla leaves a separate pattern of excitation
behind, which means the bath has gained 
information about the system, and
coherence is irreversibly lost \cite{giulini}.
Thus, another condition for AQEC is that 
excitation should be symmetrically removed
from separate funnels.
\par
The final condition comes from examining the coherence
\[
|\psi_0 \rangle |00\rangle \langle \psi_1 | \langle 00| \,
a_0^\star a_1 \times 
\] \begin{equation} 
\sum_m \left(
u_0^{(0)} c_0^{(0)}(m\Delta t) |e_0^{(0)} \rangle +
u_0^{(1)} c_0^{(1)}(m\Delta t) |e_0^{(1)} \rangle + \cdots
\right)^\dag \left(
u_1^{(0)} c_1^{(0)}(m\Delta t) |e_1^{(0)} \rangle +
u_1^{(1)} c_1^{(1)}(m\Delta t) |e_1^{(1)} \rangle + \cdots
\right)
\label{eqn:xprod} \end{equation}
The sum is the inner product of two vectors,
indexed by $m$, whose elements are 
environmental wavefunctions.   Each vector
individually has a unit norm, so by the Swartz
inequality, the sum is one if the inner product
of each element is maximum.   Thus, the final
criteria for AQEC is to have
$\sum_k u_n^{(k)} c_n^{(k)}(m\Delta t)|e_n^{(k)}\rangle$
=$\sum_k u_q^{(k)} c_q^{(k)}(m\Delta t)|e_q^{(k)}\rangle$,
for each pair of codewords $n$ and $q$, and
at each time $m\Delta t$.
Physically, we are again preventing the bath
from gaining information about the codewords.
In this case, however, the information would be
transferred by the pattern of environmental
entanglements with the bath, instead of the
excitation.
\par
This last requirement is similar
to the phase-matching requirement for
frequency mixing in non-linear optical
materials \cite{boyd}.
Consider the following contrived example. 
A qubit suffers an error, 
$\alpha |0\rangle +
\beta |1 \rangle$ $\rightarrow$
$\alpha |2\rangle + 
\beta |3\rangle$.
These four states have frequencies
$\omega_0$, $\omega_1$, $\omega_2$
and $\omega_3$, respectively.
The original coherence between $|0\rangle$
and $|1\rangle$ has been transferred
to a new set of states.  
There are no ancilla to this repair; instead,
relaxation symmetrically drives 
$|2\rangle \rightarrow |0\rangle$
and $|3\rangle \rightarrow |1\rangle$ 
at a steady rate.   Therefore, we can write 
$c_0(t) = \sqrt{\gamma} \exp(-i \{
\omega_2 t + \omega_0 (T-t) \} - \gamma t/2)$
and $c_1(t) = \sqrt{\gamma} \exp(-i 
\{ \omega_3 t + \omega_1 (T-t) \} -\gamma t/2)$.
The original coherence gains a factor of 
$\int c_0^\star(t) c_1(t) dt$ =
$\exp(i(\omega_0 - \omega_1)T) \times \gamma
/ (\gamma - i( \omega_0 - \omega_1 -
\omega_2 + \omega_3) )$.
When $\omega_0-\omega_2$ =
$\omega_1-\omega_3$,
the dynamics between the separate
funnels is indistinguishable as far as the
bath can discern, and a full repair results.
\par
To summarize, the following are sufficient
conditions for AQEC, although they may not all
be necessary.   In particular, it is likely that
the Markov assumption could be relaxed.
(\textbf{1}) The system obeys a Lindblad equation
of motion, with an evolution superoperator $\Gamma$
\cite{giulini}.   If cooling occurs on
degenerate transitions, they must be coupled to
orthogonal modes of the bath \cite{cooling,vankampen}.
(\textbf{2}) The eigenstates of $H$ consist of
codewords, $|\psi_n\rangle$, the funnel subspaces
associated with each codeword, $\{ |\phi_n^{(p)}\rangle \}$,
and the rest.   The codewords obey the conditions
of QECC \cite{knill,nielsen}, and errors transform
codewords only into their associated funnels.
If the errors are available as joint system / environmental
transforms, $U$, then check whether 
$\langle \phi_n^{(k)}|U|\psi_n\rangle$ =
$\langle \phi_q^{(k)}|U|\psi_q\rangle$ for all
$n\ne q$, and for some labeling, $k$, of the
funnel states.
(\textbf{3}) The codewords form a decoherence free
subspace with respect to the bath \cite{lidar}.
(\textbf{4}) $\Gamma$ does not transfer
amplitude between funnels.   All funnel populations
decay under $\Gamma$ into their associated
codeword populations.
(\textbf{5}) The dynamics under $\Gamma$ within
each codeword-funnel subspace are identical.
If ancilla are used, they must be excited symmetrically
between separate codewords.
The last conditions, which are the novel aspect of
this approach, are necessary in order to repair
the coherences of the codewords using dissipation.
Alternatively, one could replace the last three
conditions with criteria on the eigenvalues and
left and right eigenstates of $\Gamma$.
\section{Some Numerical Simulations of AQEC.}
\subsection{The Single Codeword Model.}
This section provides
a better idea of how AQEC works
by examining the behavior of a few 
numerical simulations.
Recall the example of keeping $S$ in
the state $|1_S\rangle$.   We now see that
we could have used 
(the states are ordered as
$|0_S 0_A \, \rangle$, 
$|0_S 1_A \, \rangle$,
$|1_S 0_A \, \rangle$ 
and $|1_S 1_A \, \rangle$):
\begin{equation}
H / \hbar = \left( \begin{array}{cccc}
\omega_{00} & 0 & 0 & \mu \\
0 & \omega_{01} & 0 & 0 \\
0 & 0 & \omega_{10} & 0 \\
\mu^\star & 0 & 0 & \omega_{11} \end{array} \right)
\;\;\;\; \mbox{instead of} \;\;\;\;
H / \hbar =
\left( \begin{array}{cccc}
r & d & 0 & 0 \\
d & 0 & 0 & r \\
0 & 0 & d & 0 \\
0 & r & 0 & d \end{array} \right).
\label{eqn:hamone} \end{equation}
As previously, $\Gamma$ has one
zero eigenvalue, $|1_S 0_A \rangle \langle 1_S 0_A|$.
However, the path by which
error correction occurs is different:
$|0_S 0_A \, \rangle 
\leftrightarrow 
|1_S 1_A \, \rangle 
\rightarrow 
|1_S 0_A \, \rangle$.
This results in a more efficient repair,
as seen by comparison of Fig.~(\ref{fig:graph2}) 
to Fig.~(\ref{fig:linent2}).
The phase of $\mu$, and the parameters
$\omega_{01}$ and $\omega_{10}$, are irrelevant, 
but as $\Delta \omega$ = $\omega_{11}-\omega_{00}$ 
increases, the first step becomes less efficient.   
However, it isn't crucial 
that $\omega_{00}$ = $\omega_{11}$ exactly.
The less optimized parameters slow down,
but do not halt, the correction process.
Note that, in contrast to QECC,
a spin-flip error at $A$ is removed 
without ever influencing $S$.
\par
\parbox{3.0in}{
\includegraphics[angle=0,width=3in,height=2.5in]
{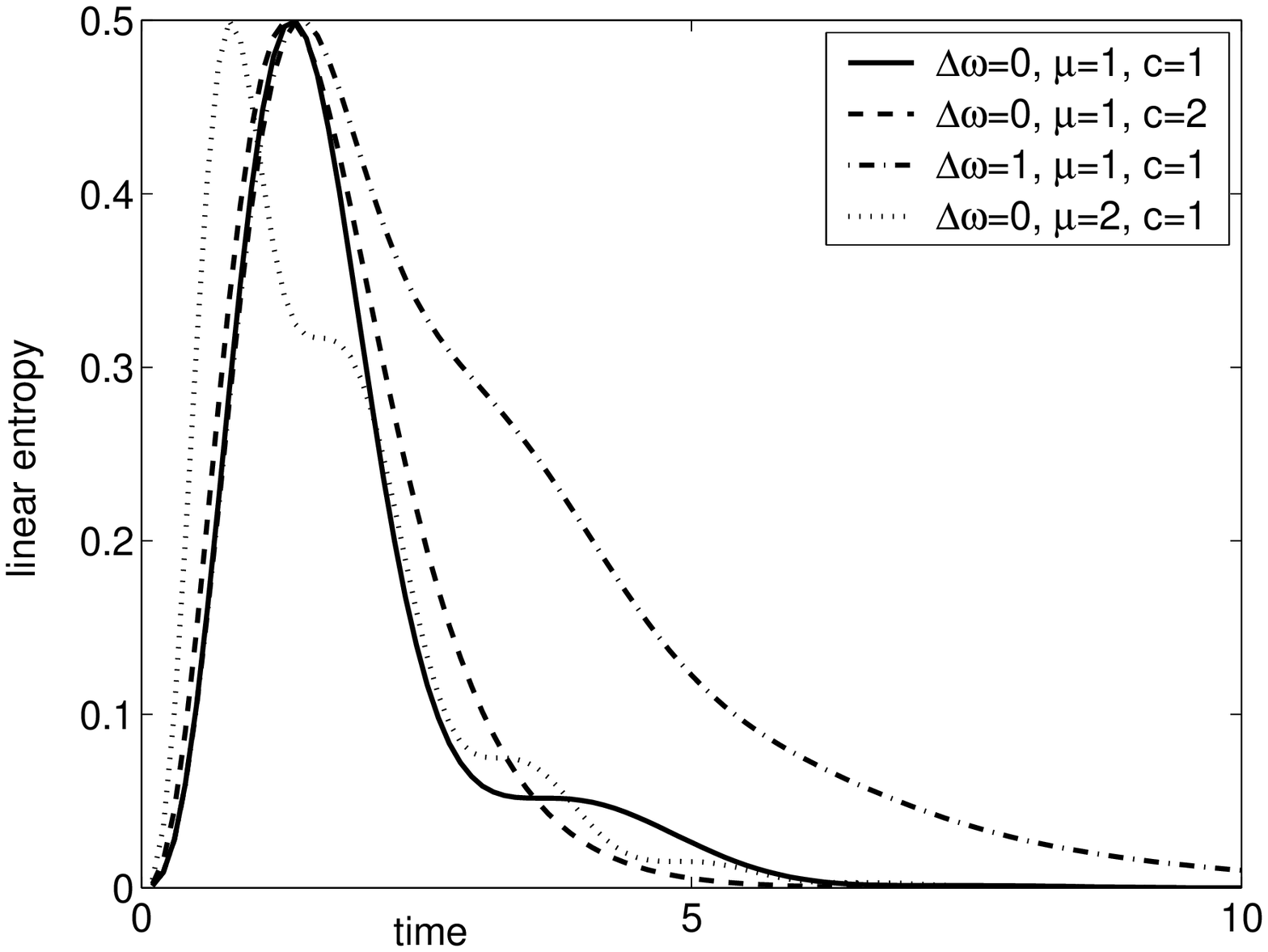}}
\parbox{3.2in}{
\refstepcounter{no_float_fig} \label{fig:linent2}
\textbf{Fig. \ref{fig:linent2}.}
The linear entropy as a function of time
during dissipative AQEC for a single stable state.
The starting state is $\rho$ =
$|0_S 0_A \rangle \langle 0_S 0_A |$.
The more rapidly that $H$ can mix the
corrupted state with an excited ancilla,
and then cool the ancilla, the more rapid
the repair.   Non-zero values of $\Delta \omega$,
or small values of $c$, lead to a slower repair.}
\par
\subsection{The Two-Codeword, Spin-flip Correcting Model.}
Let us re-examine the system that 
protects against spin-flip errors,
using three $S$ and two $A$ qubits.
The codewords are $|000,00\rangle$ and
$|111,00\rangle$.  Parameterize the
system $H$ as shown in Fig.~(\ref{fig:parms}):
\par
\parbox[b][3in][c]{3.2in}{
\[ \left( \begin{array}{c|ccc|ccc}
\omega_0 & 0 & 0 & 0 & 0 & 0 & 0 \\ \hline
0 & \omega_{e1} & \gamma_{12} & \gamma_{13} & 
\mu_{11} & \mu_{12} & \mu_{13} \\
0 & \gamma_{12}^\star & \omega_{e2} & 
\gamma_{23} & \mu_{21} & \mu_{22} & \mu_{23} \\
0 & \gamma_{13}^\star & \gamma_{23}^\star & \omega_{e3} & 
\mu_{31} & \mu_{32} & \mu_{33} \\ \hline
0 & \mu_{11}^\star & \mu_{21}^\star & \mu_{31}^\star & 
\omega_{c1} & \kappa_{12} & \kappa_{13} \\
0 & \mu_{12}^\star & \mu_{22}^\star & \mu_{32}^\star & 
\kappa_{12}^\star & \omega_{c2} & \kappa_{23} \\
0 & \mu_{13}^\star & \mu_{23}^\star & \mu_{33}^\star & 
\kappa_{13}^\star & \kappa_{23}^\star & \omega_{c3}
\end{array} \right) \;\;\; \begin{array}{c} 
|000,00\rangle \\ |001,00\rangle \\ |010,00\rangle \\ 
|100,00\rangle \\ |000,01\rangle \\ |000,10\rangle \\ 
|000,11\rangle \end{array} \]
\refstepcounter{no_float_fig} \label{fig:parms}
\textbf{Fig. \ref{fig:parms}.}
A parameterized $H$ for a
two-codeword AQEC with three $S$ and two $A$.
$H$ is block diagonal, with
the two blocks parameterized as shown
above (the lines provide a guide for the eye,
with a listing of the order of the states for
the first funnel / codeword combination).
The blocks must be identical, to within a constant
offset along the diagonal, so the dynamics between
the funnels appears indistinguishable.}
\parbox[b][3in][t]{3.0in}{
\includegraphics[angle=0,width=3in,height=3in]
{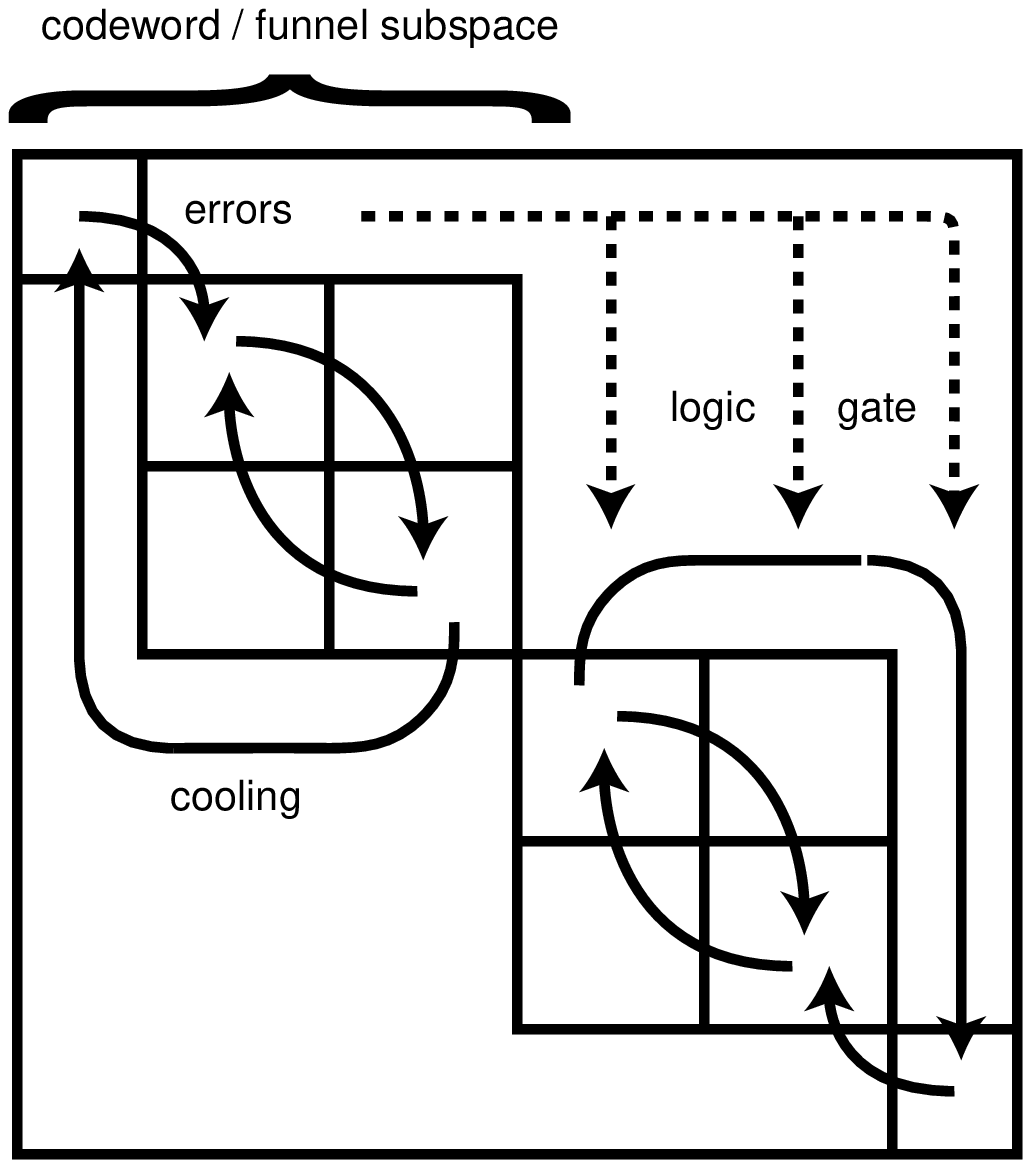} }
\par
The $\gamma$ mix the different error states,
the $\kappa$ mix the excited
ancilla states, and the $\mu$ mix
the errors with the excited ancilla states.
Previous implementations of QECC kept
the $\mu$ matrix diagonal and the 
$\gamma$, $\kappa$ = 0, so that separate errors
excited orthogonal ancilla states.
For AQEC, we must check that each of the
six eigenstates of the funnels have some non-zero
projection along a state with excited ancilla
so that population is not trapped in a funnel.
\par
Eq.~(\ref{eqn:examples}) shows some examples for $H$.
All three sets properly repair spin-flip errors, but
set (C), which is nearest in spirit to QECC, 
implements the most rapid repair.
For the simulations,
only the 14 total codeword and funnel states
are used in the numerical simulations, so
$\Gamma$ is $196 \times 196$ in size.
The matrix exponential routine of MATLAB 
\cite{matlab} was
used to produce $\exp(\Gamma \, t)$. 
The initial $\rho$ is found by tracing
the environment out from the initial
error state, $|\Psi\rangle |e_0\rangle$ +
$I_{1,x} |\Psi\rangle |e_1\rangle$ +
$I_{2,x} |\Psi\rangle |e_2\rangle$ +
$I_{3,x} |\Psi\rangle |e_3\rangle$,
where $|\Psi\rangle$ = $(1/\sqrt{2})|000,00\rangle$
+ $(\exp(i\pi/3)/\sqrt{2})|111,00\rangle$.
This state allows us to check whether 
the coherence phase is properly recovered.
In general, the environmental overlaps
$\langle e_n | e_m \rangle$
could be any complex numbers subject to
$\sum_n \langle e_n | e_n \rangle = 1$
and $|\langle e_n | e_m \rangle|^2$ $\le$ 
$\langle e_n | e_n \rangle \langle e_m | e_m \rangle$.
\begin{equation} A \;
\left( \begin{array}{c|ccc|ccc}
10 & 0 & 0 & 0 & 0 & 0 & 0 \\
\hline
0 & 2 & 1 & 0 & 1 & 0 & 0 \\
0 & 1 & 2 & 1 & 0 & 0 & 0 \\
0 & 0 & 1 & 2 & 0 & 0 & 0 \\
\hline
0 & 1 & 0 & 0 & 2 & 0 & 0 \\
0 & 0 & 0 & 0 & 0 & 2 & 0 \\ 
0 & 0 & 0 & 0 & 0 & 0 & 2
\end{array} \right)
\;\;\; B \;
\left( \begin{array}{c|ccc|ccc}
10 & 0 & 0 & 0 & 0 & 0 & 0 \\
\hline
0 & 2 & 1 & 0 & 0 & 0 & 0 \\
0 & 1 & 2 & 1 & 0 & 0 & 0 \\
0 & 0 & 1 & 2 & 1 & 0 & 0 \\
\hline
0 & 0 & 0 & 1 & 2 & 1 & 0 \\
0 & 0 & 0 & 0 & 1 & 2 & 1 \\ 
0 & 0 & 0 & 0 & 0 & 1 & 2
\end{array} \right)
\;\;\; C \;
\left( \begin{array}{c|ccc|ccc}
10 & 0 & 0 & 0 & 0 & 0 & 0 \\
\hline
0 & 2 & 0 & 0 & 1 & 0 & 0 \\
0 & 0 & 2 & 0 & 0 & 1 & 0 \\
0 & 0 & 0 & 2 & 0 & 0 & 1 \\
\hline
0 & 1 & 0 & 0 & 2 & 0 & 0 \\
0 & 0 & 1 & 0 & 0 & 2 & 0 \\ 
0 & 0 & 0 & 1 & 0 & 0 & 2
\end{array} \right)
\label{eqn:examples} \end{equation}
\par
Set Eq.~(\ref{eqn:examples},A) 
uses a single ancilla to correct 
the three independent spin-flip errors.  
Fig.~(\ref{fig:twoa5})
shows the recovery of the codeword populations 
and coherences for a spin-flip error at each $S$,
and for a spin-flip error with a set of randomly
chosen environmental overlaps, $\langle e_n | e_m \rangle$,
as given in Eq.~(\ref{eqn:overlaps}).
\begin{equation}
\left( \begin{array}{cccc}
.10 & -.7+.2i & 0 & -.3-.3i \\
& .41 & .3+.7i & .4-.2i \\
& & .27 & .8+.3i \\
& & & .22 \end{array} \right)
\label{eqn:overlaps} \end{equation}
The $H$ of Eq. (\ref{eqn:examples},A)
excites an ancilla only for the
first spin-flip error.   It repairs the other spin-flip
error by mixing all the errors together.   The numerical
simulations shown in Fig.~(\ref{fig:twoa5}) show this
process in detail.   Note that if the all three spin-flips
entangle the system with the environment, then 
the linear entropy of the system is initially non-zero.
\par
\parbox{3.0in}{
\includegraphics[angle=0,width=3.0in,height=2.5in]{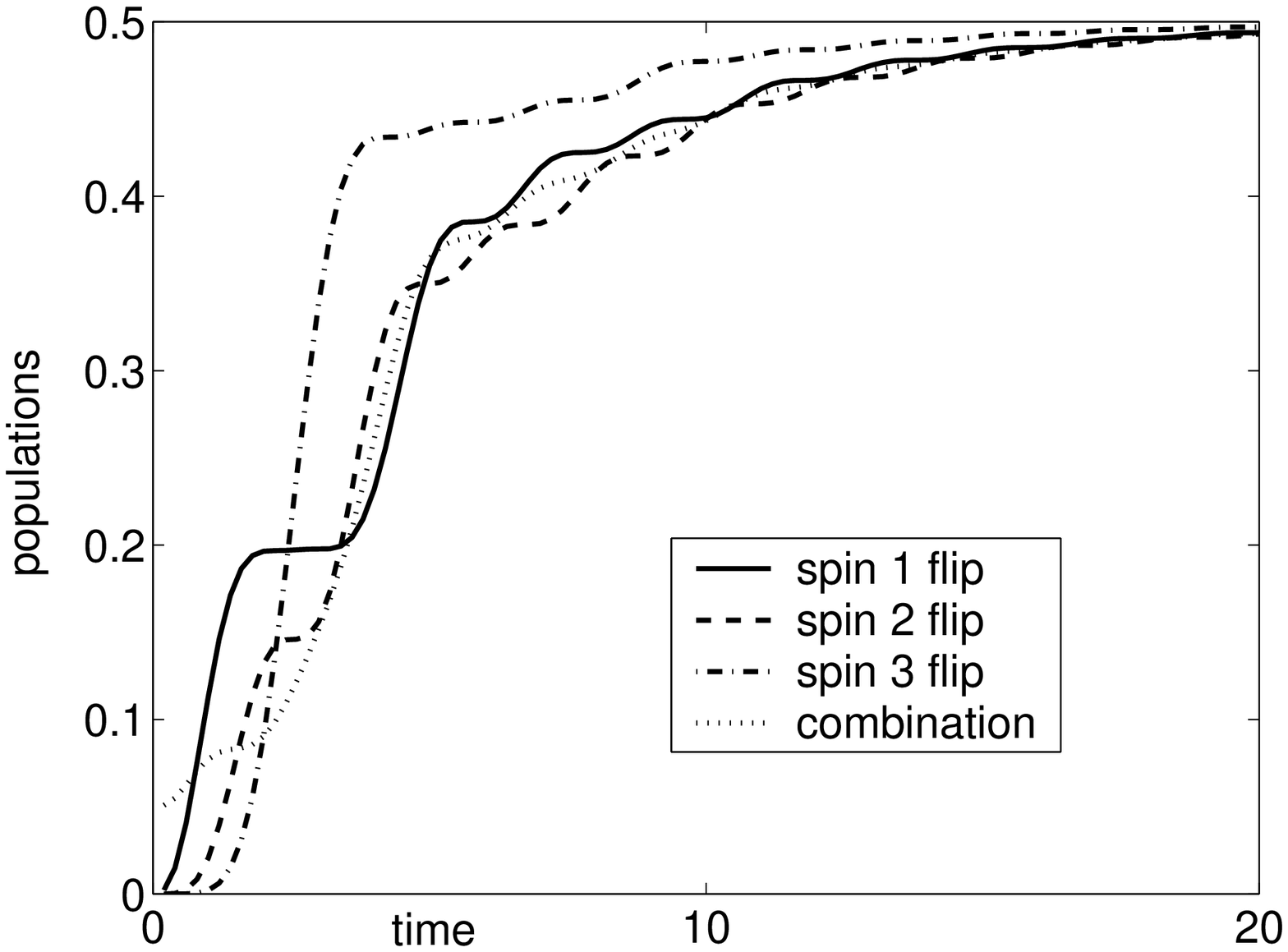}
} \hfill
\parbox{3.0in}{
\includegraphics[angle=0,width=3.0in,height=2.5in]{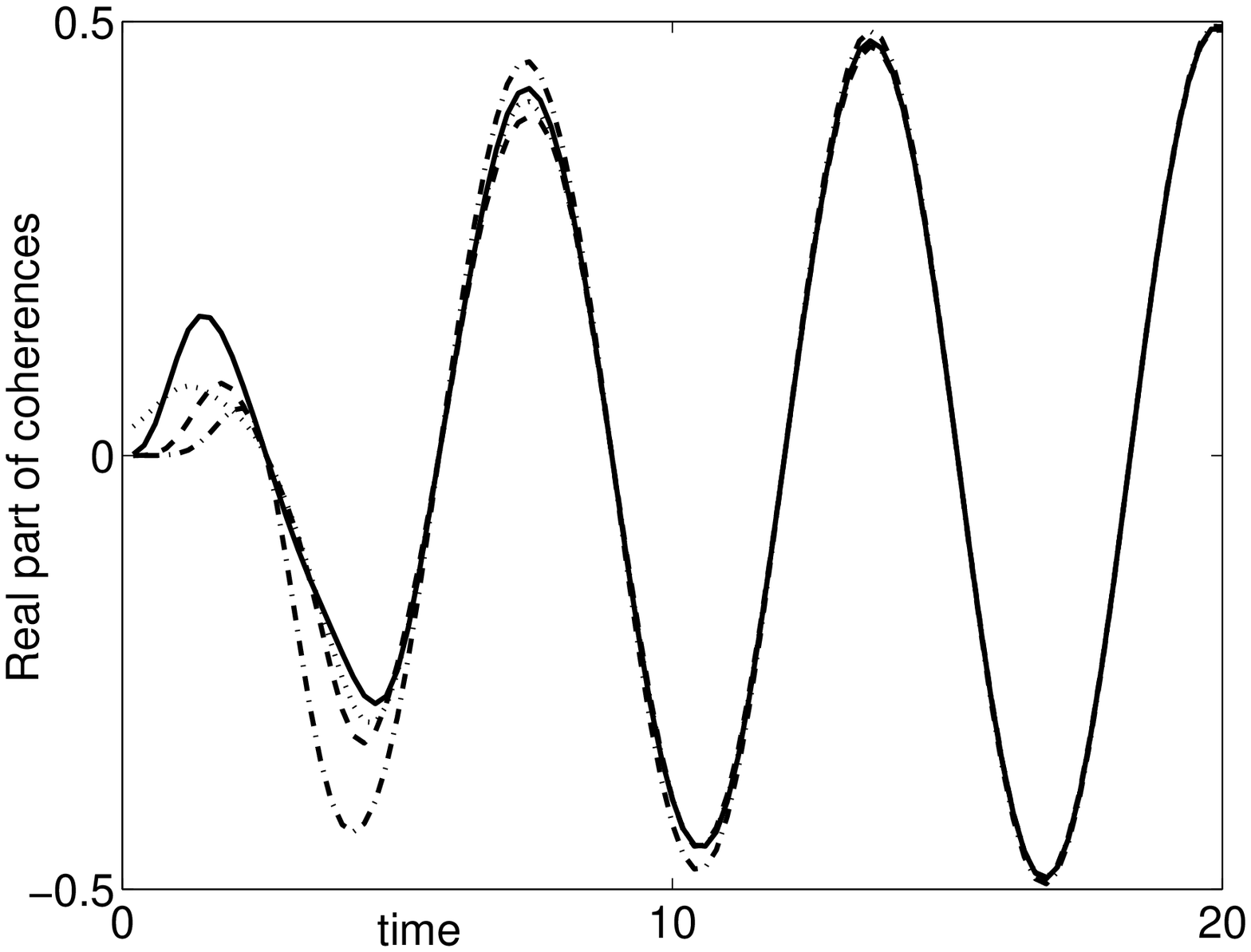}
} \par
\parbox{3.0in}{
\includegraphics[angle=0,width=3.0in,height=2.5in]{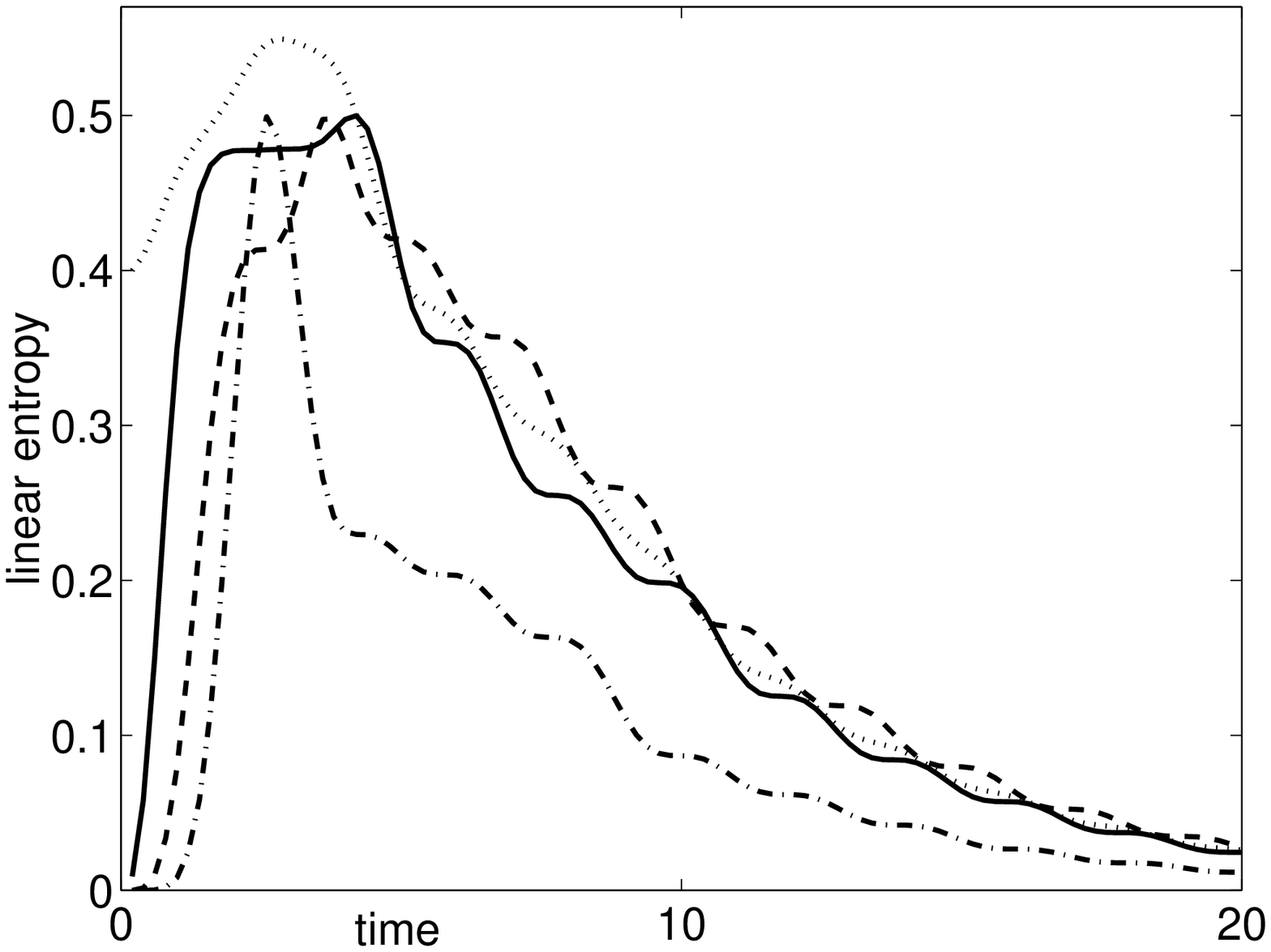}
} \hfill
\parbox{3.0in}{ 
\refstepcounter{no_float_fig} \label{fig:twoa5}
\textbf{Fig. \ref{fig:twoa5}.}
Repair of the populations (upper left),
coherences (upper right), and the linear
entropy (lower), for errors in which each
$S$ alone flips, or there is a correlated
flip of all the spins whose environmental
overlaps are given by Eq. (\ref{eqn:overlaps}).
All three graphs have the same legend.
The repair process is a linear dynamics, so 
it also repairs correlated single spin-flips.
Populations and coherences of both codewords
follow identical paths.
The parameter set is Eq. (\ref{eqn:examples},A),
but with the second codeword / funnel 
offset by $\Delta \omega$ = 1
in order to show the oscillation of the
coherences.   The cooling rates were
$c_1, c_2$ = 1.} \par
Thus, AQEC can expell the information about
which error occurred at the same time as the
error is repaired.  Eq.~(\ref{eqn:examples},B) 
is another example of this.
It mixes together all the errors, and all the
excited ancilla states.   Because it does so symmetrically
between the separate codewords, the errors are repaired.
Fig.~(\ref{fig:twoa1}) shows the populations 
for all the funnel states after the first spin is flipped.
It can be observed that all the states are transiently
excited:  the state vector ``swirls around'' in each
funnel as the repair occurs.
\par
\parbox{3.0in}{
\includegraphics[angle=0,width=3in,height=2.5in]
{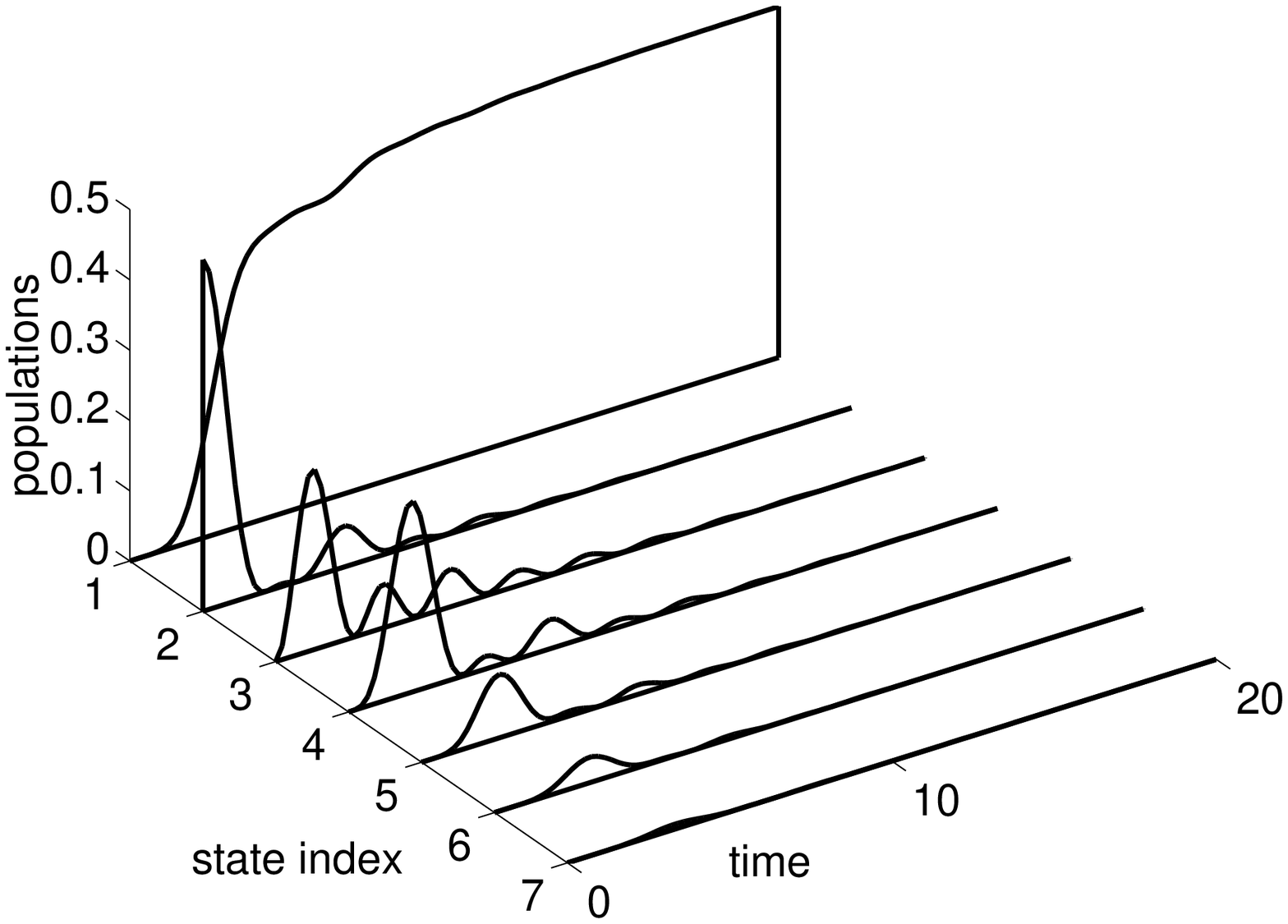}} \hfill
\parbox{3.0in}{
\refstepcounter{no_float_fig} \label{fig:twoa1}
\textbf{Fig. \ref{fig:twoa1}.}
The populations of the codewords (at back),
of the three different spin-flip states 
(middle three), and of the three orthogonal
excited ancilla states (forward three),
during a repair after the first spin is flipped.
The parameter set is Eq. (\ref{eqn:examples},B),
with cooling rates $c_1, c_2$ = 1.
It is permissible under AQEC to mix together
different errors during the repair, so long
as the dynamics between the separate codeword / funnel
subspaces is indistinguishable to the bath.}
\par
What happens when condition (\textbf{5}) is violated?
There are two possibilities:  excite the ancilla
asymmetrically between the codewords, or have different
dynamics between the two codeword / funnel subspaces.
Fig.~(\ref{fig:twoa2}) shows the first case, for which
Eq.~(\ref{eqn:examples},A) was used, but modified for
the funnel surrounding $|111,00\rangle$ by setting
$\mu_{11}=\mu_{12}=1/\sqrt{2}$.   The cooling rates
were $c_1, c_2$ = 1, and the error was $I_{1,x}$.
The partially orthogonal
ancilla states do not allow $\Gamma$ to return
$\rho$ to a pure state.   In fact, if $\mu_{11}$ were
set to zero for the second codeword, then there would be
no element in $\Gamma$ to transfer 
$|000,10\rangle \langle 111,01|$ to
$|000,00\rangle \langle 111,00|$.  
The bath gains information about the system through
excitation.   The coherence asymptotically approaches
$0.3530 \exp(i\pi(0.3333)\,)$, with correct phase
but low magnitude.
\par
\parbox{3.0in}{
\includegraphics[angle=0,width=3in,height=2.5in]
{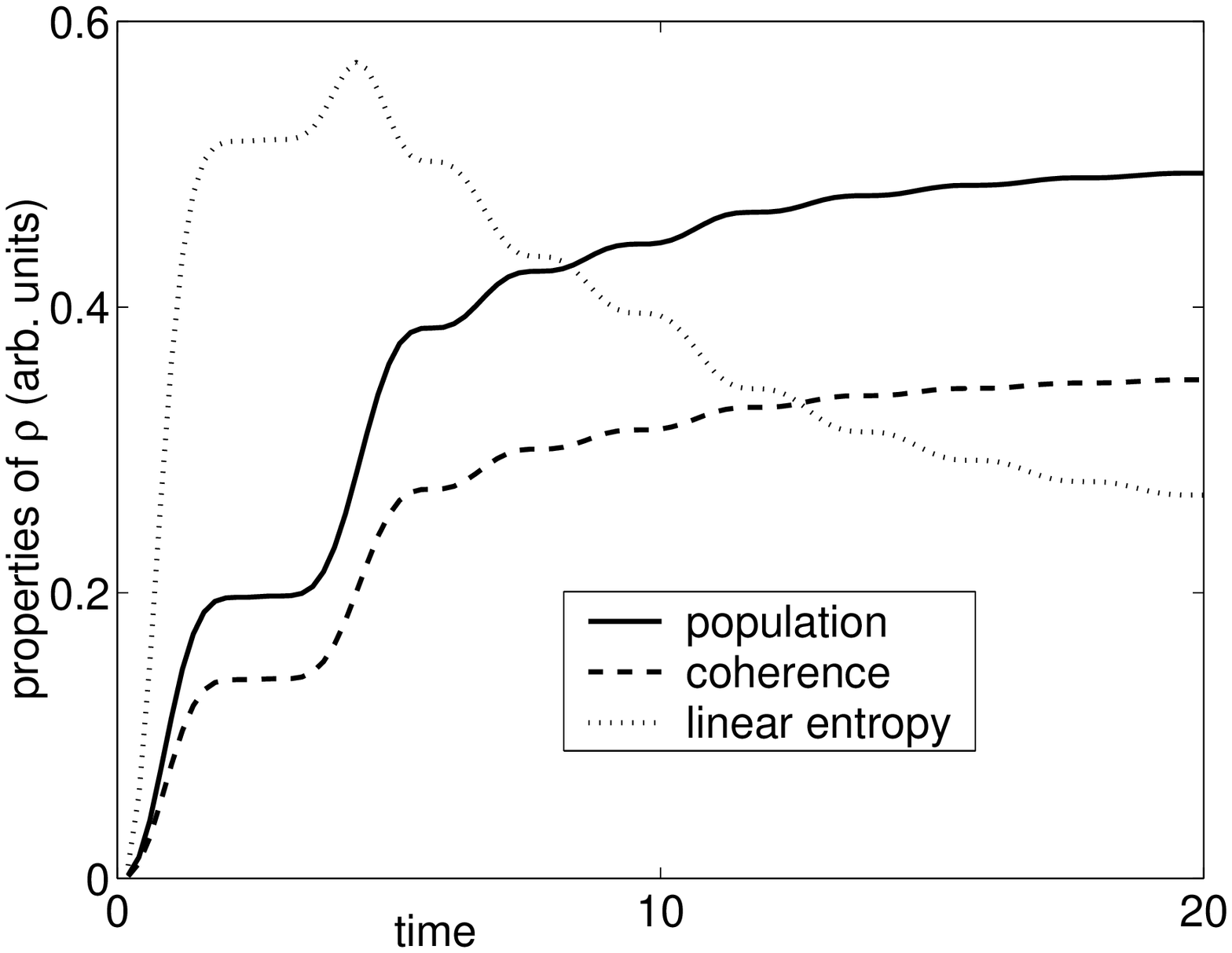}} \hfill
\parbox{3.0in}{
\includegraphics[angle=0,width=3in,height=2.5in]
{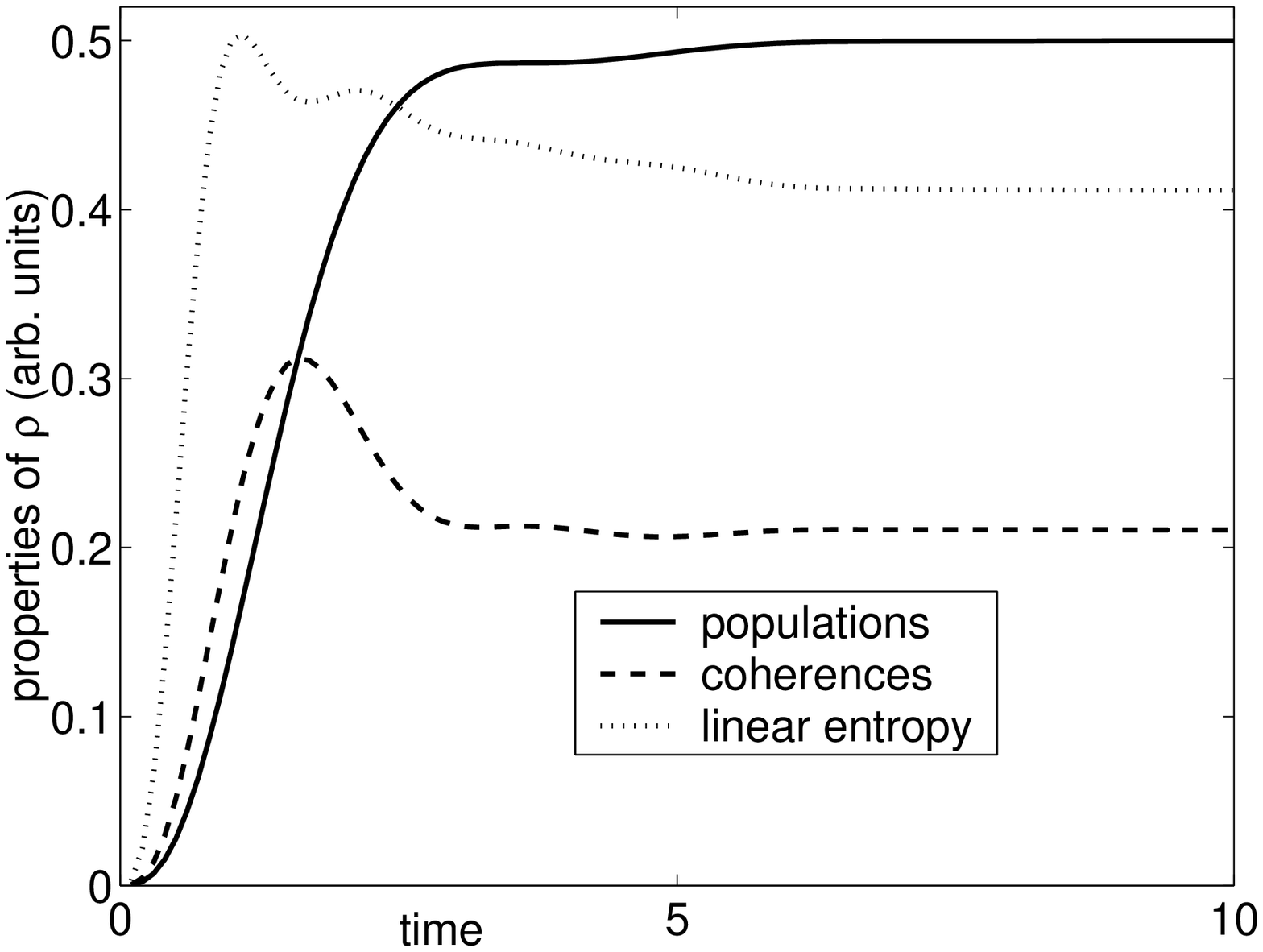} }
\par
\parbox{3.0in}{ 
\refstepcounter{no_float_fig} \label{fig:twoa2}
\textbf{Fig. \ref{fig:twoa2}.}
A partial repair
due to the use of partially orthogonal
excited ancilla states, 
$|10\rangle$ and $(|10\rangle+|01\rangle)/\sqrt{2}$,
between the two codewords.
The populations (solid line) are repaired,
but the coherence magnitudes are not (dashed line).}
\hfill
\parbox{3.0in}{
\refstepcounter{no_float_fig} \label{fig:twoa3}
\textbf{Fig. \ref{fig:twoa3}.}
A partial repair resulting 
from a dissimilar dynamics between the two
codeword / funnel subspaces.   The populations
are repaired (solid line), but the coherence
magnitudes are not (dashed lines).}
\par
\vspace{0.1in}
The second possibility is shown in Fig.~(\ref{fig:twoa3}).
Here, the ancilla are excited symmetrically,
but the dynamics between the funnels
is not equivalent, and coherence is again lost.
The parameter set is Eq.~(\ref{eqn:examples},C),
but $\mu_{11}$=2 for the second funnel,
so the mixing was more rapid.
The cooling rates were $c_1, c_2$ = 1, and the
error was $I_{1,x}$.   Again, the population
is repaired, but some coherence is lost.
\section{A Proposed Test System.}
We now give a physically realizable example
of an AQEC system that implements Shor's three-qubit,
majority code against spin-flip errors \cite{shor}.
It can not repair phase-flip errors, so it is not
suitable for a quantum computer.   It serves to
illustrate a method by which to find systems
suitable for AQEC.   It is also encouraging that a
system can be found without making recourse to exotic
interactions.
\par
Our strategy is to find a system that obeys multiple
conservation laws that the errors violate.
Consider a system with an observable $A$
such that $[H,A]=0$, and an error $E$ where $[A,E]=E$.
The simultaneous eigenstates of $H$ and $A$,
$|\epsilon,a\rangle$, have the property that
$A E |a\rangle$ = $(a+1)E|a\rangle$.
Thus, choosing codewords with $a$ = 0 and 2
gives rise to funnel states with $a$ = 1 and 3.
If, in addition, there is a unitary $B$ such that
$B|\epsilon,a\rangle = |\epsilon,a+2\rangle$,
then the funnel states can be mapped onto one
another, and their dynamics are equivalent.
\par
Let three spin 1/2 particles be lined up along
the $z$ axis, in a zero static magnetic field.
They interact by point dipolar
$D_{nm}(I_{n,x}I_{m,x}+I_{n,y}I_{m,y}-2I_{n,z}I_{m,z})$
and exchange
$J_{nm}(I_{n,x}I_{m,x}+I_{n,y}I_{m,y}+I_{n,z}I_{m,z})$
terms \cite{slichter}.
Dipolar interactions decrease with distance
as $r^{-3}$, so for equally spaced spins,
$D_{12}$=$D_{23}$=$8D_{13}$=$\zeta$,
where $\zeta$ can be as large as 0.1 cm$^{-1}$
\cite{esr}.
\par
Assuming the dipolar interactions dominate,
the level diagram of the spins is given in
Fig.~(\ref{fig:spins},A).   The spins attempt to
mutually align, resulting in ground states
of $|000\rangle$ and $|111\rangle$.   These
are the codewords.   A spin-flip error, $I_{n,x}$,
is equivalent to rotating a spin by $\pi$
about the $x$ axis.   Since this requires work
against the dipolar field, dissipation can
repair these errors.   The funnels come from
the conservation of the spin angular momentum
about the $z$ axis, $\sum_n I_{n,z}$, which
has eigenvalues denoted as $m_z$.   An error
changes $m_z \rightarrow m_z \pm 1$.
The codewords have $m_z = \pm 3/2$.
The funnel surrounding $|000\rangle$
(levels A-C of Fig.~(\ref{fig:spins},A) )
has $m_z = -1/2$, and the funnel
surrounding $|111\rangle$ has $m_z = +1/2$
(levels D-F).
\par
\vspace{0.2in}
\parbox{4.0in}{
\includegraphics[angle=0,width=4in,height=2.0in]
{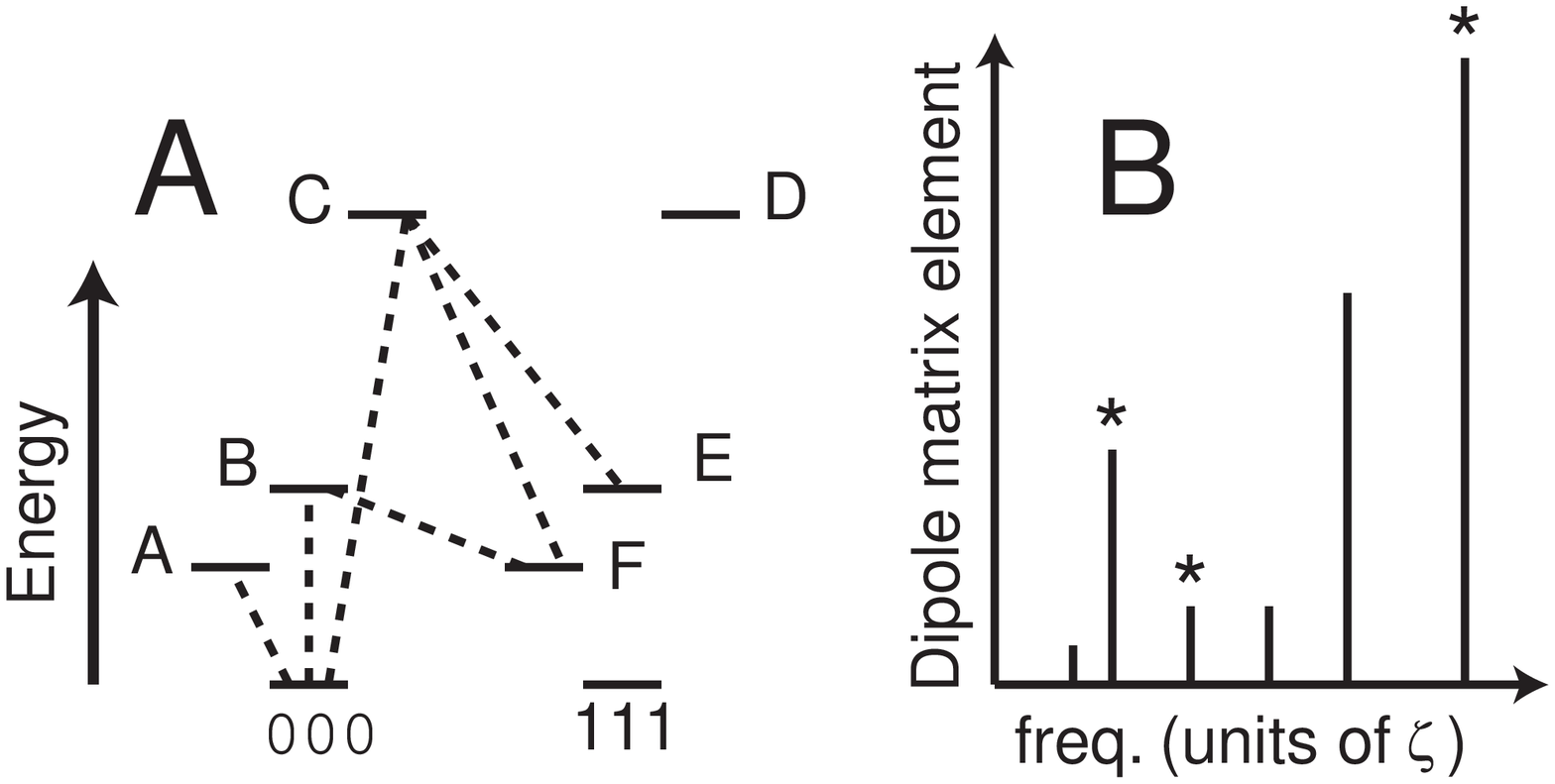} } \hfill
\parbox{2.0in}{ 
\refstepcounter{no_float_fig} \label{fig:spins}
\textbf{Fig. \ref{fig:spins}.}
(\textbf{A}) The level diagram for the three spin system.
The ground states are the codewords.   States
A-C with $m_z=-1/2$ form the funnel for $|000\rangle$.
The dashed lines show the dipole-allowed transitions 
for the first funnel, with symmetric transitions
for the second funnel.
(\textbf{B}) The spectrum of dipole-allowed transitions.
The starred lines represent the funnel to codeword
transitions that should be cooled.}
\par
\vspace{0.2in}
Spontaneous emission of a photon with an $x$ polarized
$B$ field will symmetrically de-excite the degenerate
funnels.   Thus, we could use photons as the ancilla
for this system.   There are several advantages to this
approach.   First, it is difficult to selectively
cool a single spin, but it is easy to cool an
electromagnetic resonator.   Second, the validity
of the Markov approximation for a damped resonator mode
is better understood \cite{scully,walls}.
A $y$-polarized $B$ field will anti-symmetrically
de-excite the funnels, so an error $I_{n,y}$ will be
``repaired'' to the phase-flip error $I_{n,z}$.
This phase-flip error can not be repaired, because
it requires no work to rotate a spin about the $z$
axis for this system.
\par
There are two last difficulties to overcome.
The dashed lines of Fig.~(\ref{fig:spins},A) show
the dipole-allowed transitions for the $m_z = -1/2$
states.   Their strengths are proportional to matrix
elements of the operator $\sum_n I_{n,x}$.
Point dipolar interactions alone are not sufficient
for AQEC, because $\langle B|\sum_n I_{n,x}|000\rangle$
is always zero, no matter how the spins are positioned
along the $z$ axis.   This unwanted symmetry is broken
by setting $J_{12}$ = $J_{13}$ = 0 and $J_{23}=0.2\zeta$.
Actually, any $0 < |J_{23}| \le 0.5\zeta$ will work.
Using the above parameters gives rise to the spectrum
of Fig.~(\ref{fig:spins},B).   The second difficulty
is that, besides spontaneous emission of $y$-polarized
$B$-field photons, there are also transitions between
the funnels.   We wish to cool only the starred transitions
(at $0.64\zeta$, $1.03\zeta$, and $2.39\zeta$), but
not the un-starred ones 
(at $0.39\zeta$, $1.36\zeta$ and $1.75\zeta$).
\par
This can be achieved by placing the spins at the
center of a resonator whose modes are only resonant
with the starred transitions.
Consider a rectangular, conducting cavity
of linear dimensions $a$, $b$, and $d$.
Resonances are indexed as transverse electric
(TE$_{mnp}$, where $n+m>0$, $p > 0$) and transverse magnetic 
(TM$_{mnp}$, where $n,m>0$, $p \ge 0 $) modes \cite{gandhi},
with frequencies $\omega_{mnp}$ =
$\sqrt{(m/a)^2+(n/b)^2+(p/d)^2}$ in units 
of cm$^{-1}$ if the cavity lengths are in cm.
Each mode produces either a linearly
polarized electric or magnetic field, 
or no field, at the center.
One can invert the above to find that a
resonator with dimensions $a$=2.32/$\zeta$, 
$b$=0.87/$\zeta$, and $d$=4.28/$\zeta$,
has TE$_{102}$, TE$_{104}$, and TE$_{122}$
modes resonant with the starred transitions.
Each mode has an $x$ polarized $B$ field at the
spins.   Another 29 modes exist with $\omega < 2.5\zeta$.
Of those that produce $B$ fields at the spins,
the nearest to a funnel-funnel transition is TE$_{302}$, 
which is offset by 0.018$\zeta$ from the C-E
transition.   A resonator Q $\gg 76$ is required
to suppress emission of this transition.
For $\zeta \approx 0.1$ cm$^{-1}$, microwave
resonators can achieve this goal.  This larger 
$\zeta$ is also desirable because the cold bath 
must satisfy $T \ll (hc/k)\zeta \approx$
0.1 K, a not outrageous requirement.
\section{Discussion}
AQEC borrows the same structure for storing information
as in QECC, but implements the error correction in a
different way.   In NMR terminology, the qubit of the
above system is hidden in the triple quantum coherence
of the spins.   The novel aspect is that dissipation can
be used to directly repair not only the codeword populations,
but also the coherences.   The criteria for this is
simply summed up by demanding that excitation, and environmental
entanglements, be expelled from the codewords in a symmetric
manner.
\par
Especially interesting is the possibility that an AQEC qubit
exists that can protect against both spin- and phase-flip errors.
Exchange interactions may prove more useful in this regard.
Being isotropic interactions, they resist the rotation of a
spin about any axis.   One difficulty with using only exchange
interactions, is that dipole-allowed transitions vanish, so a
symmetric de-excitation of all the funnels by photons becomes
problematic.   Another open question is how AQEC behaves when
it is scaled up to large numbers of codewords.   It is, however,
helpful to contemplate an error correction scheme that requires
no additional burden to the programmer.
\section{Conclusions}
Conditions are given by which the dissipative evolution of a
system, coupled  to a cold Markovian bath, can be used to
implement automatic quantum error correction.   The new condition,
necessary to repair codeword coherences, requires a symmetric
de-excitation of separate codewords, and an equivalent
dynamics between the different codeword / funnel subspaces.
They resemble the conditions of phase-matching in nonlinear
optics.   A test case, that of Shor's majority-code against
spin-flip errors \cite{shor}, is proposed.   It utilizes
well known dipolar and exchange interactions between spins,
and dipole-allowed transitions with the modes of a resonator.
\subsection{Acknowledgements}
We gratefully acknowledge support from
the Air Force Office of Scientific Research.
\end{document}